\documentclass[preprint]{aastex63}

\usepackage{multirow}
\usepackage{booktabs}

 \submitjournal{ApJL}

\graphicspath{{./}{figures/}}
\begin{document}

\title{First report of a solar energetic particle event observed by China$'$s Tianwen-1 mission in transit to Mars}

\correspondingauthor{Xiaoping Zhang, Cunhui Li, Gang Li, Shuwen Tang}
\email{xpzhangnju@gmail.com, licunhui@spacechina.com,
gang.li@uah.edu, tangsw@impcas.ac.cn}

\author[0000-0003-4245-3107]{Shuai Fu}
\altaffiliation{These authors contributed equally.}
\affiliation{State Key Laboratory of Lunar and Planetary Sciences, Macau University of Science and Technology, Taipa 999078, Macau, China}

\author[0000-0002-9829-3811]{Zheyi Ding}
\altaffiliation{These authors contributed equally.}
\affiliation{School of Geophysics and Information Technology, China University of Geosciences (Beijing), Beijing 100083, China}

\author{Yongjie Zhang}
\altaffiliation{These authors contributed equally.}
\affiliation{Institute of Modern Physics, Chinese Academy of Sciences, Lanzhou 730000, China}

\author[0000-0002-4306-5213]{Xiaoping Zhang}
\affiliation{State Key Laboratory of Lunar and Planetary Sciences, Macau University of Science and Technology, Taipa 999078, Macau, China}

\author[0000-0003-2689-9387]{Cunhui Li}
\affiliation{Science and Technology on Vacuum Technology and Physics Laboratory, Lanzhou Institute of Physics, Lanzhou 730000, China}
\affiliation{School of Earth and Space Sciences, University of Science and Technology of China, Hefei 230026, China}

\author[0000-0003-4695-8866]{Gang Li}
\affiliation{Department of Space Science and CSPAR, University of Alabama in Huntsville, Huntsville, AL 35805, USA}

\author{Shuwen Tang}
\affiliation{Institute of Modern Physics, Chinese Academy of Sciences, Lanzhou 730000, China}

\author{Haiyan Zhang}
\affiliation{Science and Technology on Vacuum Technology and Physics Laboratory, Lanzhou Institute of Physics, Lanzhou 730000, China}

\author[0000-0001-8894-525X]{Yi Xu}
\affiliation{State Key Laboratory of Lunar and Planetary Sciences, Macau University of Science and Technology, Taipa 999078, Macau, China}

\author[0000-0002-8887-3919]{Yuming Wang}
\affiliation{School of Earth and Space Sciences, University of Science and Technology of China, Hefei 230026, China}

\author[0000-0002-8707-076X]{Jingnan Guo}
\affiliation{School of Earth and Space Sciences, University of Science and Technology of China, Hefei 230026, China}

\author[0000-0002-4299-0490]{Lingling Zhao}
\affiliation{Department of Space Science and CSPAR, University of Alabama in Huntsville, Huntsville, AL 35805, USA}

\author{Yi Wang}
\affiliation{Science and Technology on Vacuum Technology and Physics Laboratory, Lanzhou Institute of Physics, Lanzhou 730000, China}

\author{Xiangyu Hu}
\affiliation{Science and Technology on Vacuum Technology and Physics Laboratory, Lanzhou Institute of Physics, Lanzhou 730000, China}

\author[0000-0002-1066-2273]{Pengwei Luo}
\affiliation{State Key Laboratory of Lunar and Planetary Sciences, Macau University of Science and Technology, Taipa 999078, Macau, China}

\author[0000-0002-7667-3178]{Zhiyu Sun}
\affiliation{Institute of Modern Physics, Chinese Academy of Sciences, Lanzhou 730000, China}

\author{Yuhong Yu}
\affiliation{Institute of Modern Physics, Chinese Academy of Sciences, Lanzhou 730000, China}

\author[0000-0001-9635-4644]{Lianghai Xie}
\affiliation{State Key Laboratory of Space Weather, National Space Science Center, Chinese Academy of Sciences, Beijing 100190, China}

\begin{abstract}
Solar energetic particles (SEPs) associated with flares and/or coronal mass ejection (CME)-driven shocks can impose acute radiation hazards to space explorations. To measure energetic particles in near-Mars space, the Mars Energetic Particle Analyzer (MEPA) instrument onboard China$'$s Tianwen-1 (TW-1) mission was designed. Here, we report the first MEPA measurements of the widespread SEP event occurring on 29 November 2020 when TW-1 was in transit to Mars. This event occurred when TW-1 and Earth were magnetically well connected, known as the Hohmann-Parker effect, thus offering a rare opportunity to understand the underlying particle acceleration and transport process. Measurements from TW-1 and near-Earth spacecraft show similar double-power-law spectra and a radial dependence of the SEP peak intensities. Moreover, the decay phases of the time-intensity profiles at different locations clearly show the reservoir effect. We conclude that the double-power-law spectrum is likely generated at the acceleration site, and that a small but finite cross-field diffusion is crucial to understand the formation of the SEP reservoir phenomenon. These results provide insight into particle acceleration and transport associated with CME-driven shocks, which may contribute to the improvement of relevant physical models.
\end{abstract}

\section{Introduction}
\label{sect1}
Mars, also known as the Red Planet, is the fourth planet from the Sun at an average radial distance of about 1.52 astronomical units (AU) and the second smallest planet in the solar system after Mercury (e.g., \citealp{stinner2005journey}). Exploration of Mars could provide clues to those questions concerning the origin and evolution of life, and in the long term, the Red Planet could one day become a destination for human existence. But before that, several tricky issues should be considered, one of which is the inevitable radiation exposure associated with energetic charged particles from space \citep{zeitlin2004overview}. \textcolor{black}{As we know, the radiation environment of Mars is much more severe than that of Earth due to the lack of a global magnetic field and a thick enough atmosphere, which
has been a huge obstacle to scientific exploration of this planet (e.g., \citealp{hassler2014mars,Guo+etal+2021}).}

Space radiation environment is a complex field consisting of charged particles with energies spanning several orders of magnitude \citep{nelson2016space}. The main sources of these particles are Galactic cosmic rays (GCRs), which are highly energetic and highly penetrating particles \citep{simonsen2020nasa}. Another important source of space radiation is solar energetic particles (SEPs). SEPs are \textcolor{black}{produced by solar flares and/or coronal mass ejections (CMEs)-driven shocks, and they may dominate on short time scales (usually several hours to days) and thus pose acute radiation hazards to space \citep{ehresmann2018energetic}. In contrast, GCRs are a chronic and steady background of energetic charged particles in the heliosphere, while SEPs are transient, sporadic, and unpredictable bursts that can occur at any stage of a solar cycle, but mostly around solar maximum (e.g., \citealp{zeitlin2013a,hu2022extreme}).}
%Despite the respective roles of solar flares and CME-driven shocks during SEP events are not fully understood, the latter is extensively considered to be the most productive and largest accelerator of SEPs in the inner heliosphere (e.g., \citealp{li2021b,Ding2022,hu2022extreme}).
In large SEP events, proton fluxes can be several orders of magnitude higher than the background fluxes of GCRs  \citep{Seedhouse2018}.
%and cause severe radiation hazards to hardware and astronauts in space
For the purpose of better serving future Mars missions, a comprehensive understanding of the space radiation environment near Mars is urgently needed, particularly for large SEP events.

%SEP events are the injection of energetic charged particles from the Sun into interplanetary space.
%There are two distinct types of SEPs: impulsive and gradual SEPs \citep{cane1986two,reames1999particle}. Specifically, impulsive SEP events come from solar flares and are associated with short duration of soft X-ray emission, narrow longitudinal spread, type III radio bursts, and enhanced ratios of $^3$He/$^4$He \citep{kallenrode2003current}, while gradual SEP events usually originate from CME-driven shocks, characterized by long duration of soft X-ray emission, wide longitudinal spread, type II radio bursts, and non-measurable enhanced ratios of $^3$He/$^4$He \citep{reames2013two}.
\textcolor{black}{A central task in studying SEP events is to understand their temporal and spatial evolution in the heliosphere.} It is found that particle intensity (also termed differential flux) profiles observed by different spacecraft with large latitudinal, longitudinal, and radial separations usually evolve to a similar level in the late phase of large SEP events, as first noticed by \citet{mckibben1972azimuthal} and later referred to as the reservoir phenomenon by \citet{roelof1992low}. \textcolor{black}{To date, the underlying cause of this phenomenon has not been fully understood, but }a plausible explanation is that those accelerated particles can be trapped behind the CME as the magnetic bottle for an extended period \citep{roelof1992low}. \citet{wang2015simulations} argued that many factors, such as parallel diffusion, adiabatic cooling, source intensity, and perpendicular diffusion, jointly contribute to the reservoir phenomenon. Furthermore,  the radial dependence of SEP intensities (and fluences) is also an interesting topic in SEP studies. In previous works, the radial dependence of SEP peak intensities was usually characterized by a functional form of $J_{max}=kR^{\alpha}$, where $R$ is the radial distance of a spacecraft from the Sun,  $J_{max}$ is the peak intensities of SEPs, and $\alpha$ is the radial dependence index. Regarding the power law index $\alpha$, there are some debates in different research works. For example, \citet{kallenrode1997statistical} derived an index $\alpha$ of -5.5 to -4.5 based on the analysis of 4--13 MeV protons in 44 SEP events, while \citet{lario2006radial} studied the 4--13 MeV and 27--37 MeV proton intensities in 72 SEP events and found $\alpha$ in the range of -2.7 to -1.9.

%Prior studies have been greatly constrained by the limited observational data, especially the relatively scarce measurements beyond 1 AU.

To better understand the radiation environment of energetic particles near Mars, the China National Space Administration (CNSA) launched the Mars Energetic Particle Analyzer (MEPA) instrument in July 2020 to probe the energy spectra, fluxes, and elemental compositions of energetic charged particles. This instrument is on board China$'$s first Mars exploration mission, Tianwen-1 (TW-1), which is a fairly comprehensive mission incorporating orbiting, landing, and roving in one mission, and is composed of an orbiter and a lander/rover \citep{wan2020china}. The highly capable and configurable TW-1/MEPA instrument is able to detect multiple species of energetic charged particles over a relatively wide energy range, specifically, 0.1--12 MeV electrons, 2--100 MeV protons, 25--300 MeV alpha particles, and heavy ions up to iron. \textcolor{black}{For more details on this instrument, see
references \citet{tang2020calibration} and \citet{li2021design}.}

\textcolor{black}{On 29 November 2020, a widespread SEP event erupted and was observed by multiple widely separated spacecraft within 1 AU, including the Advanced Composition Explorer (ACE), \textcolor{black}{Geostationary
Operational Environmental Satellite (GOES),} Parker Solar Probe (PSP), Solar Orbiter (SolO), Solar and Heliospheric Observatory (SOHO), and Solar-Terrestrial Relations Observatory (STEREO), but remains unreported beyond 1 AU (e.g., \citealp{kollhoff2021first}).} During the event, TW-1 was at about 1.39 AU, and the onboard TW-1/MEPA was fortuitous and happened to be switched on to fully capture the particle intensity variations of this event. More coincidentally, the Earth and TW-1 were tied on the same field line at that time (see the discussion below), which was named the Hohmann Parker effect by \citet{posner2013hohmann}, providing a unique opportunity to understand the underlying physics of particle transport along a magnetic field line with different interplanetary magnetic field (IMF) path lengths. In this work, we performed a comparative study of energetic proton observations between TW-1 ($\sim$1.39 AU) and near-Earth ($\sim$1 AU) missions from 29 November to 5 December 2020, with implications for better knowledge of the underlying particle acceleration and transport processes in the inner heliosphere and for possible future human missions to Mars.

\section{DATASET AND METHOD}
\label{sect2}
\subsection{In situ proton measurements}
We collected proton flux measurements from two sources: one from the TW-1/MEPA instrument, and the other one from three near-Earth spacecraft, including ACE, SOHO, and Wind. These near-Earth spacecraft station at the Sun-Earth first Lagrange ($L$1) point, which is about 1.5 million kilometers from the Earth ($\sim$0.01 AU), along the Sun-Earth connection line.

The TW-1/MEPA data were acquired from the Planet Exploration Program Scientific Data Release System (\url{http://202.106.152.98:8081/marsdata/web/datainfo/main.action\#}). The level-2 ACE/EPAM/LEMS120 data were available from the ACE Science Center (\url{http://www.srl.caltech.edu/ACE/ASC/level2/}). The level-2 SOHO/ERNE data were available for download at \url{https://srl.utu.fi/erne_data/}. The level-3 SOHO/COSTEP/EPHIN data were available from \url{http://ulysses.physik.uni-kiel.de/costep/level3/l3i/}. The Wind/EPACT/STEP data were downloaded from \url{https://spdf.gsfc.nasa.gov/pub/data/wind/epact/step/differential-ion-flux-1hr/}. All data were acquired from 28 November to 5 December 2020, and were collated to a 1-hour resolution.

\subsection{Band-function form fit}
Considering the measured SEP proton spectra show a clear double power-law feature, we fit them using a Band-function form, given by \citet{band1993batse},
\begin{equation}
F(E)=\left\{\begin{array}{lll}
C E^{-\gamma_{a}} \exp \left(-E / E_{0}\right) & \text { for } & E<\left(\gamma_{b}-\gamma_{a}\right) E_{0}, \\
C \textcolor{black}{E^{-\gamma_{b}}}\left[\left(\gamma_{b}-\gamma_{a}\right) E_{0}\right]^{\left(\gamma_{b}-\gamma_{a}\right)} \exp \left(\gamma_{a}-\gamma_{b}\right) & \text { for } & E>\left(\gamma_{b}-\gamma_{a}\right) E_{0},
\end{array}\right.
\label{eq1}
\end{equation}
where $F(E)$ is the particle fluence, $C$ is a normalization constant, $E_0$ is the spectral break energy in units of MeV/nuc, and $\gamma_a$ and $\gamma_b$ are the spectral indexes in the low and high energy end, respectively.

\subsection{Radial and IMF path length dependence of SEP peak intensities}
We use Equation (\ref{eq2}) to calculate the index of the peak intensity radial dependence ($\alpha$) and Equation (\ref{eq3}) to calculate the index of the peak intensity IMF path length dependence ($\beta$):
\begin{equation}
J(E)_{\max } \propto R^{\alpha(E)}, \quad \alpha(E)=\log _{\frac{R_{1}}{R_{2}}} \frac{J(E)_{\max,1}}{J(E)_{\max,2}},
\label{eq2}
\end{equation}

\begin{equation}
J(E)_{\max } \propto L^{\beta(E)}, \quad \beta(E)=\log _{\frac{L_{1}}{L_{2}}} \frac{J(E)_{\max , 1}}{J(E)_{\max , 2}},
\label{eq3}
\end{equation}
where $J(E)_{max}$ is the measured proton peak intensity in the energy interval $E$, $R$ ($L$) is the radial distance (IMF path length) of a spacecraft from the Sun, with the subscript 1 for near-Earth spacecraft ($R_1$ = 1.0 AU, $L_1$ = 1.152 AU), and the subscript 2 for at TW-1 ($R_2$ = 1.39 AU, $L_2$ = 1.818 AU).

\begin{figure}[ht!]
\epsscale{1.1}
\plotone{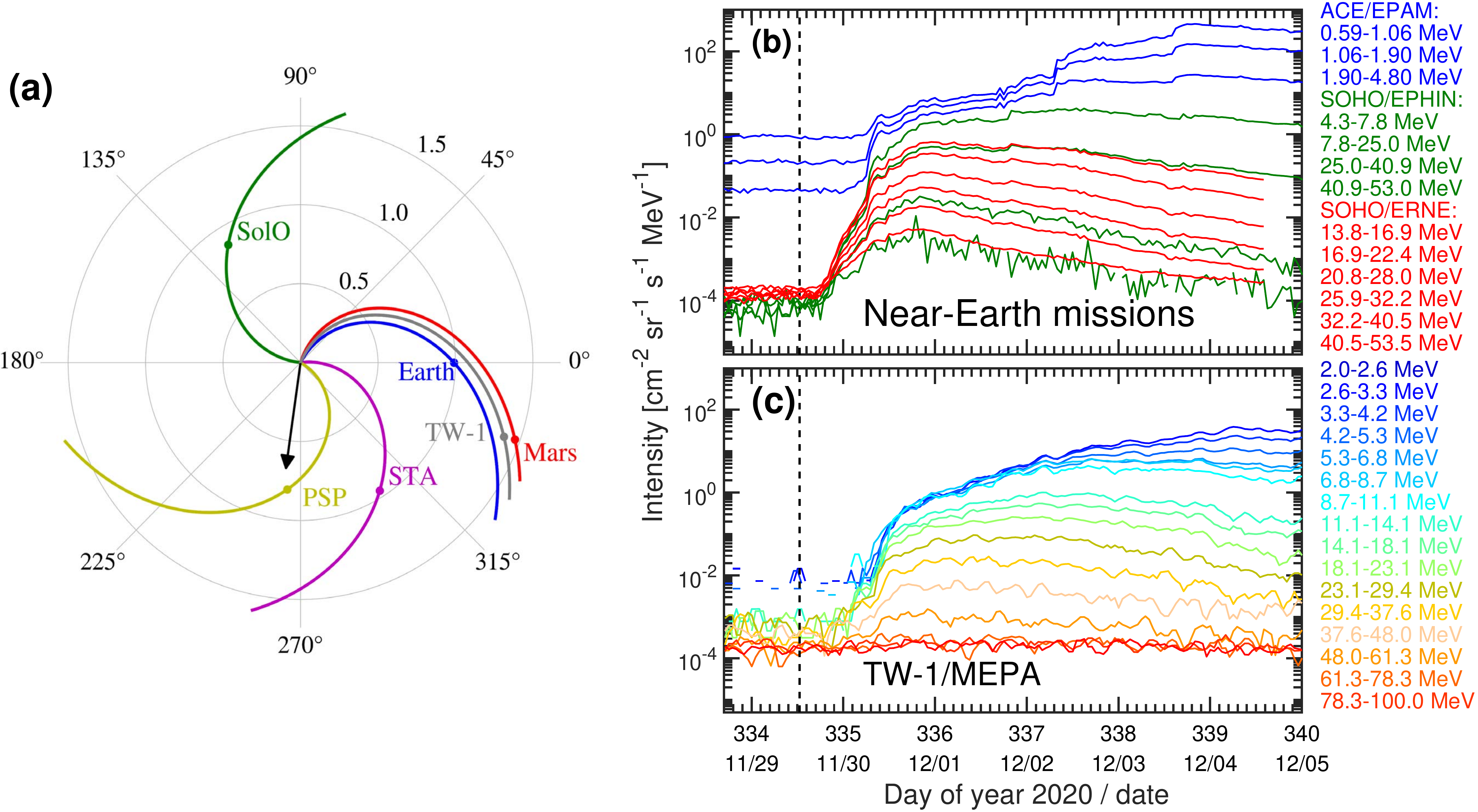}
   \caption{Locations of TW-1 (grey point), Mars (red point), Earth (blue point), STA (purple point), PSP (yellow point), and SolO (green point) at 12:00 UT on 29 November 2020 (a). The black arrow indicates the location of the active region associated with the event. The hourly averaged proton time-intensity profiles as measured by near-Earth missions (b) and TW-1/MEPA (c). The vertical dashed line indicates the onset of the flare (12:34 UT on 29 November 2020).}  \label{Fig1}
\end{figure}

\section{Results}
\label{sect3}
\subsection{Overview of the SEP event on 29 November 2020}
\label{subsect3.1}
The 29 November 2020 SEP event originated from a fast and relatively wide CME, associated with an M4.4 class X-ray flare from the active region (AR) 12790. Figure \ref{Fig1}(a) illustrates the relative locations of multiple spacecraft with large radial and longitudinal separations in heliocentric Earth ecliptic (HEE) coordinate at 12:00 UT (Universal Time, same as below) on 29 November 2020. The TW-1 spacecraft was located at 20 degrees east of the Earth and about 1.39 AU from the Sun, and some other locations, including Earth, Mars, PSP, SolO, and STEREO-Ahead (STA) are also shown. \textcolor{black}{The AR (solar flare) was located at E98 seen from the Earth \citep{kollhoff2021first,Kouloumvakos2022A&A}, as indicated by the black arrow.} The nominal Parker spiral IMF lines connecting each location are shown by solid colored curves. The spacecraft locations and the magnetic field footpoints within 1 AU are taken from \citet{kollhoff2021first}. The longitudinal separation between the AR and PSP is $\sim$2 degrees, between the AR and STA is $\sim$51 degrees. From in-situ magnetic field measurements, the CME-driven shock reached PSP (at 0.81 AU from the Sun) at 18:35 UT on 30 November 2020 and reached STA at 07:23 UT on 1 December 2020. Moreover, this event was also observed by SolO, TW-1, and near-Earth spacecraft. Of note here is that the magnetic field line connecting the Earth is calculated by using the 12-hour averaged solar wind speed ($\sim$391 km/s) measured before the onset of the flare. Since TW-1/MINPA (Mars Ion and Neutral Particle Analyzer) was not operating during that period, there were no direct solar wind observations at TW-1. However, assuming that the solar wind stream observed at the Earth does not vary significantly within two days prior to arriving at the Earth, then it is easy to see that the observations at the Earth and at the TW-1/MINPA correspond to plasma from the same footpoint. This is because the travel time of a solar wind parcel with the speed of 391 km/s for 0.39 AU is 1.73 days. For a solar rotation of 14.2 degrees per day, this translates to 24.45 degrees. Since Mars is 20 degrees east to the Earth, Mars and Earth are in very good magnetic connection (assuming a Parker field geometry) in this event. This conclusion is also supported by the recent works of \citet{zhang2022and} and \citet{fan2022solar} in which the TW-1/MINPA data was available a few days before this event (between 20 and 27 November 2020). With this solar wind speed, the longitudinal angle between the flare and the magnetic footpoint of the Earth is $\sim$157 degrees, and it is $\sim$162 degrees for TW-1. In the work of \citet{moradi2019propagation} and \citet{bian2021stochastic}, it was shown that the angular spreading of magnetic field lines at 1 AU due to footpoint random walk on the source surface can be as large as 10 degrees. \textcolor{black}{Other relevant work on magnetic field line spreading and lengthening from turbulence has also been done by \citet{Ragot2006ApJ} and \citet{Laitinen2019ApJ}}. Furthermore, as discussed in \citet{li2021b}, the intrinsic uncertainty associated with a CME-driven shock implies that observers separated longitudinally within $\sim$10 degrees at 1AU can be effectively regarded as connected to the same part of the shock. In this work, we therefore assume that the Earth and TW-1 are on the same Parker magnetic field line and are connected to the same area at the shock front. This assumption is well supported by \citet{posner2013hohmann}, who found that a spacecraft travelling to Mars would first \textcolor{black}{maintain} a good magnetic connectivity to Earth and then to Mars, the so-called Hohmann-Parker effect.

Figures \ref{Fig1}(b) and \ref{Fig1}(c) plot the time-intensity profiles as observed by near-Earth spacecraft and TW-1 from day 334 (29 November) to 340 (5 December) of the year 2020, respectively. The vertical dashed line indicates the onset of the flare at 12:34 UT on November 29. Figure \ref{Fig1}(b) shows the ensemble measurements from the ACE/EPAM (0.59--4.8 MeV), SOHO/EPHIN (4.3--53.0 MeV), and SOHO/ERNE (13.8--53.5 MeV), suggesting that the event is a mid-sized SEP event with a maximum proton energy around 60 MeV. The Earth was located west of the source and therefore observed a gradual increase of low-energy (e.g., $<$5 MeV) proton intensities. In comparison, high-energy protons (e.g., $>$30 MeV) show a delayed enhancement (several hours after the onset of flare) and peaked around day 335.8. The peak is followed by a long-duration decay. The onset time for the highest-energy bin of 40--53 MeV is about 6.6 hours after the onset of the flare. This can be explained by an extended shock acceleration process and a varying magnetic connection to the CME-driven shock (or the lack of it) for an observer \citep{ding2020modeling,Ding2022,li2021b}. \textcolor{black}{High-energy protons are only accelerated when the CME-driven shock is still close to the Sun \citep[e.g.][]{li2003energetic, li2005acceleration, hu2017modeling, ding2020modeling, li2021b, 2022ApJ...930...51L}}, while low-energy protons can be accelerated at shock front over a wide range of longitudes and at large distances from the Sun. Note that the magnetic footpoint of Earth is approximately 157 degrees west of the flare. \textcolor{black}{
Since the flare was located at E98 degrees, even if we assume a half-width of the shock to be $\sim$98 degrees, the Earth cannot connect to the shock until the shock arrives at and passes 1 AU.} However, there is no measurable plasma signature of shock arrival from near-Earth spacecraft. Therefore, particles observed at the Earth early on must propagate to the Earth-connection field lines by cross-field diffusion. Since the earliest high-energy particles take at least 6 hours to reach the Earth, this can be used to put an upper limit of the strength of cross-field diffusion as protons propagate from the Sun to the Earth. Comprehensive modeling studies as
done in \citet{li2021b} are needed to understand the role of perpendicular diffusion in this event. Figure \ref{Fig1}(c) plots the proton time-intensity profiles at energies of 2--100 MeV measured by TW-1/MEPA. Obviously, the observed profiles show similar enhancements and decay phases compared to the measurements near the Earth, which further supports the assumption that the Earth and TW-1 have similar magnetic connections to the shock. Therefore, in the following analysis, we assume the Earth and TW-1 are tied on the same field line, enabling us to investigate the physics of particle transport along a magnetic field line with different path lengths and ignore the first-order effect of an inhomogeneous particle source at the shock front.

\begin{figure}[ht!]
\epsscale{1.1}
\plotone{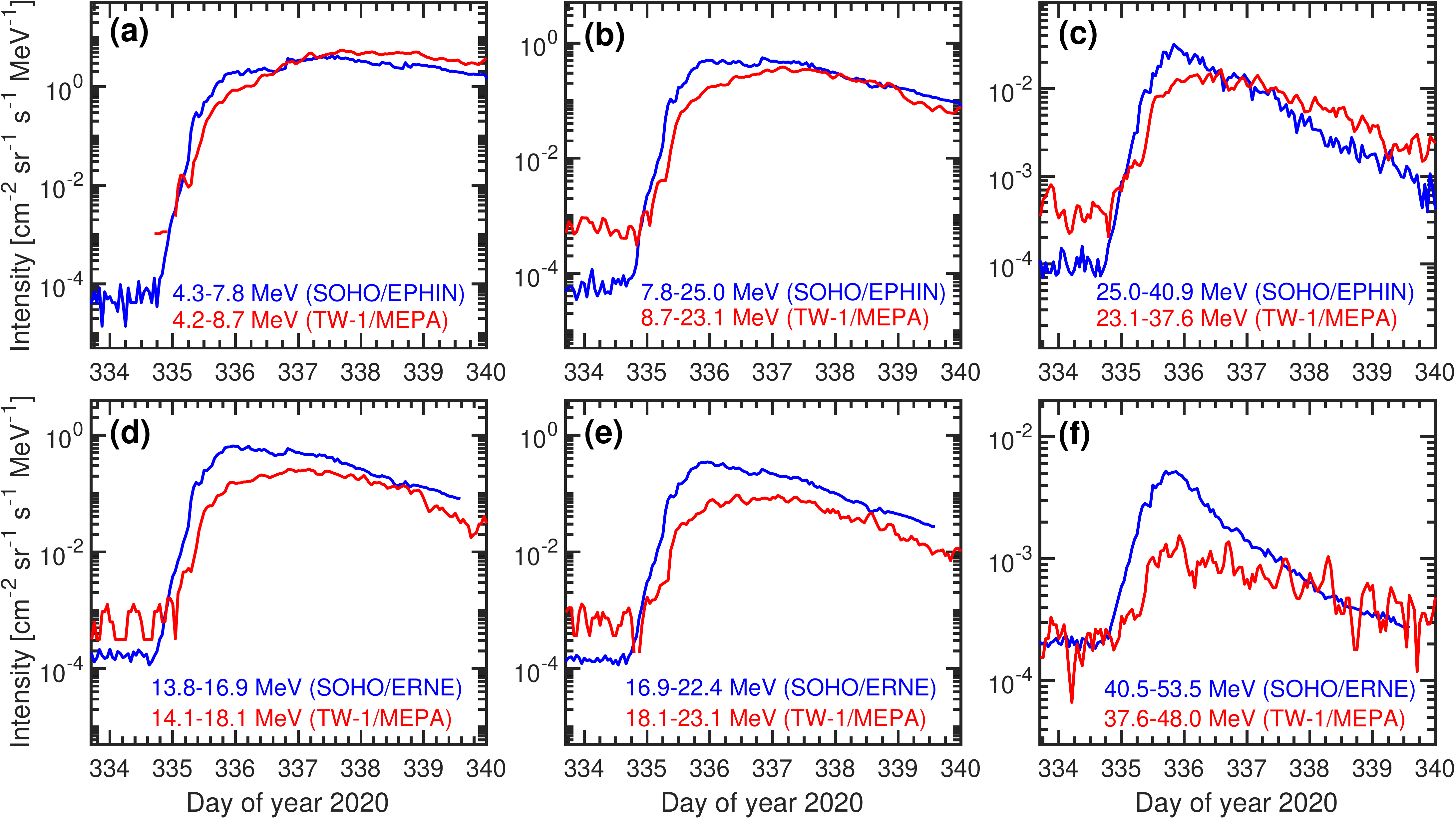}
   \caption{Comparisons of the observed proton time-intensity profiles between TW-1/MEPA and SOHO/EPHIN (a--c), and between TW-1/MEPA and SOHO/ERNE (d--f), at similar energies. The observations of TW-1 are in red, and the observations near the Earth are in blue.}  \label{Fig2}
\end{figure}

\subsection{Comparison of proton time-intensity profiles}
\label{subsect3.2}
In Figure \ref{Fig2}, we compare the proton time-intensity profiles in similar energy bins from the near-Earth measurements and TW-1/MEPA. The upper (lower) row represents the comparison between the SOHO/EPHIN (SOHO/ERNE) and TW-1/MEPA. The corresponding energy channels are labeled in each panel, ranging from $\sim$4 MeV to $\sim$50 MeV. It is worth noting that within 48 hours after the onset of the associated flare, the proton intensities of TW-1 are always lower than those around the Earth. Under the assumption that parallel diffusion coefficient is significantly larger than cross-field diffusion coefficient, we can envision the following scenario: at the beginning of this event, some accelerated protons first propagate to Earth-connected field line via cross-field diffusion and these particles then propagate to the Earth and TW-1 along the same field line. Comparing observations at the Earth and TW-1/MEPA requires a good understanding of several transport processes, including convection, pitch-angle scattering, and adiabatic deceleration. These processes combine to yield the observed radial dependence of the SEP peak intensities, which will be discussed in more detail below. After day 337 (2 December), the decay phase of SEPs from TW-1 observation is similar to that of Earth for multiple energy bins. \textcolor{black}{This can be explained by the so-called reservoir phenomenon \citep{roelof1992low,Reames1997ApJ}}, which argues that in the late phase of large SEP events, the SEPs are nearly uniformly distributed throughout the inner heliosphere. Recent studies suggested that magnetic mirroring and the effect of perpendicular diffusion are important in the reservoir phenomenon \citep{wang2021statistical}.

Although cross-field diffusion coefficient is significantly smaller than parallel diffusion coefficient, given enough time, energetic particles can still occupy a broad range of longitudes through cross-field diffusion. In fact, in this event, since the Earth and TW-1 are not magnetically connected to the shock front, cross-field diffusion is necessary to understand these observations. We speculate that cross-field diffusion plays a key role in forming the reservoir of SEPs in this event.
\textcolor{black}{We note that the SEP event of 29 November 2020 bears an interesting resemblance to the 5 December 2006 event: Both events have a fast CME accompanying a large east limb flare and the onset of SEPs in both events are delayed hours after the flare. In the 2006 SEP event, the onset of SEPs was preceded by a small energetic neutral atom (ENA) event \citep{mewaldt2009} in the energy range of a few MeVs. Since for eastern event, an earlier signal due to ENA can be clearly discerned from the SEPs, it was curious to examine if there were also ENA precursors in this event. Assuming a 2 (4) MeV ENA particle, the travel time from the Sun to Mars is $\sim$ 3.25 (2.3) hours. Upon further examinations, however, we do not find ENA precursors in this event.
}

\subsection{Comparison of proton fluence spectra}
\label{subsect3.3}
\begin{figure}[ht!]
\epsscale{1.12}
\plotone{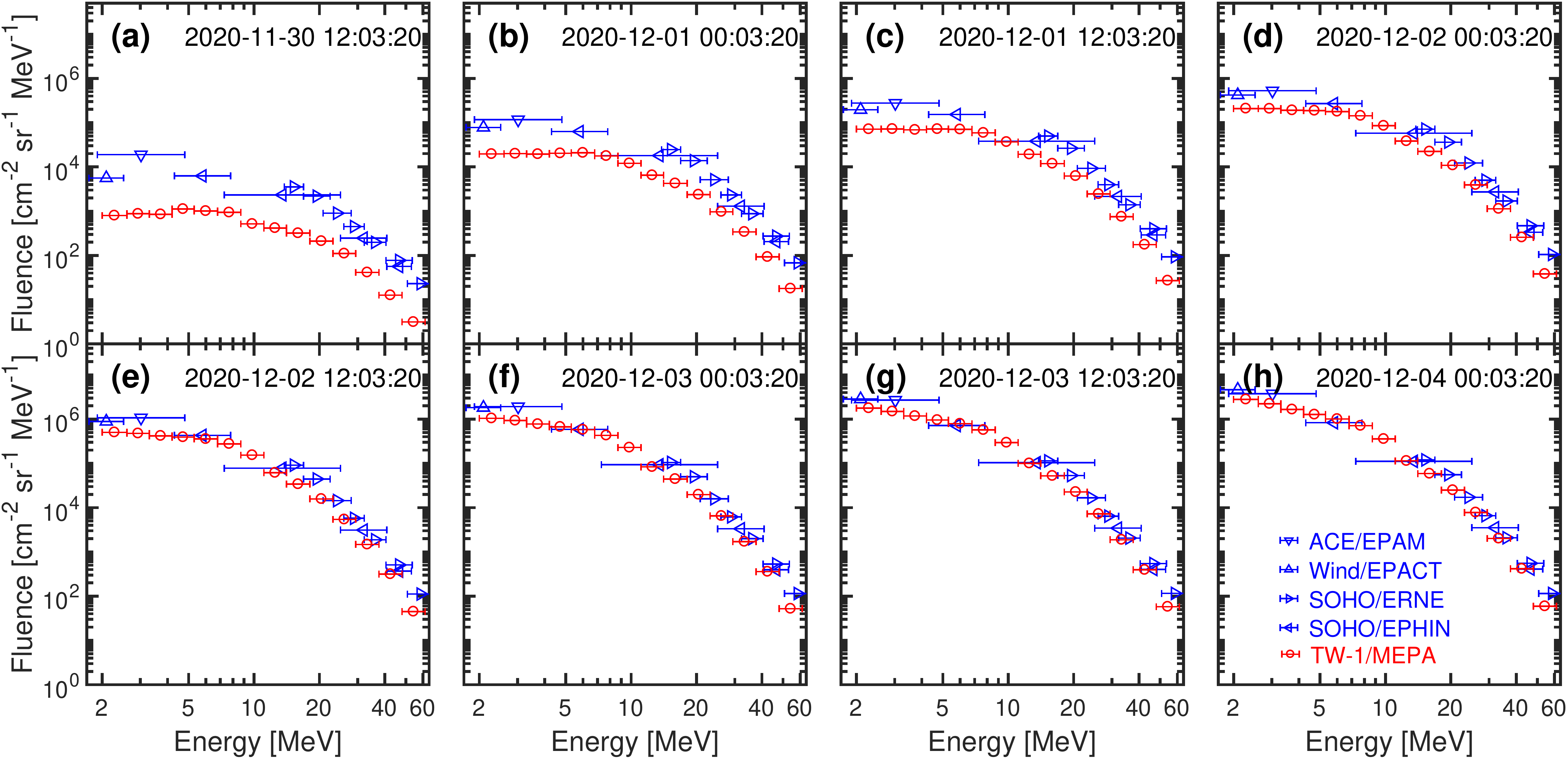}
   \caption{Comparisons of the proton fluence spectra with a 12-hour interval increase between TW-1/MEPA (red points) and near-Earth spacecraft (blue points). The starting time for these spectra is 2020-11-30 00:03:20 UT, and the ending times are marked in each panel. The horizontal bar represents the range of each energy interval.}  \label{Fig3}
\end{figure}

\textcolor{black}{Figure \ref{Fig3} plots the temporal evolution of proton fluence spectra with a 12-hour interval increase, with the starting time of 2020-11-30 00:03:20 UT and the ending time marked on the top right of each panel. To match the energy range of TW-1/MEPA, the spectra at the Earth use data from the measurements of ACE/EPAM, Wind/EPACT, SOHO/EPHIN, and SOHO/ERNE. Figure \ref{Fig3}a shows that the low-energy fluence of TW-1 is clearly lower than that of the near-Earth spacecraft in the first 24 hours, reflecting that the larger the IMF path length, the more significant the transport effect. As time increases, however, the two time-integrated proton spectra gradually approach each other (Figures \ref{Fig3}b and c), and they almost overlap 60 hours and onward after the solar flare erupted (Figures \ref{Fig3}d-h).}

The event-integrated spectra show a clear double power-law feature, prompting us to utilize a commonly used Band-function form \citep{band1993batse} to fit the spectrum in Figure \ref{Fig3}(h), which is integrated for 108 hours after the onset of the flare. Based on Equation \ref{eq1}, the spectrum near the Earth has a spectral index ($\gamma_a$) $\sim$1.34 for the low-energy end, $\sim$6.36 ($\gamma_b$) for the high-energy end, and a break energy ($E_0$) of $\sim$8.1 MeV. The corresponding values of TW-1 are $\sim$0.67, $\sim$6.92, and $\sim$5.5 MeV, respectively. The smaller spectral index at the low-energy end of TW-1 can be understood in terms of transport effects where pitch-angle scattering and adiabatic deceleration are dominant factors for the propagation of low-energy particles. A radial-dependent low-energy spectral index is an important feature in understanding the mean free path of low-energy particles. At high energies, the spectral indexes for both the Earth and TW-1 are similar and they are softer than those from most large SEP events which are typically in the range of 2$\sim$5 \citep{mewaldt2012energy}. Furthermore, the break energies at the Earth and at the TW-1 are very close, suggesting that this break and the spectrum above the break show no clear radial dependence. Some earlier work suggested that a double power law or a power law with an exponential decay at high energies is the consequence of a finite lifetime and a finite size of shock \citep{forman1983time,ellison1985shock}. \citet{mason2012interplanetary} studied the interplanetary transport effects in some large SEP events using a transport model that includes magnetic focusing, convection, adiabatic deceleration, and pitch angle scattering. They found that the spectral break location differs slightly through the transport and the particle spectrum at high energy shows a small radial dependence. They suggested that the spectral break must be a consequence of the acceleration process in the source region. In contrast, by fine-tuning solar wind turbulence levels, \citet{li2015scatter} and \citet{zhao2017effects} have argued that a double power spectrum can result from a single power-law spectrum at the source, but gradually developing as particles propagate out from the Sun. In this scenario, the break location, as well as the spectral shape above the break location, depends on the radial distance of the observer. In this work, the similarity of the spectra between Earth and TW-1 indicates that transport effects play a minor role in causing the spectral break. The observations at the Earth and Mars suggest that the double-power-law spectrum is likely generated at the acceleration site. Furthermore, the very soft high-energy spectral index may suggest that the shock acceleration efficiency is not strong in this event. This is consistent with the maximum proton energy of around 60 MeV in this event.

\subsection{Radial and IMF path length dependence of peak intensities}
\label{subsect3.4}
\begin{figure}[ht!]
\epsscale{0.85}
\plotone{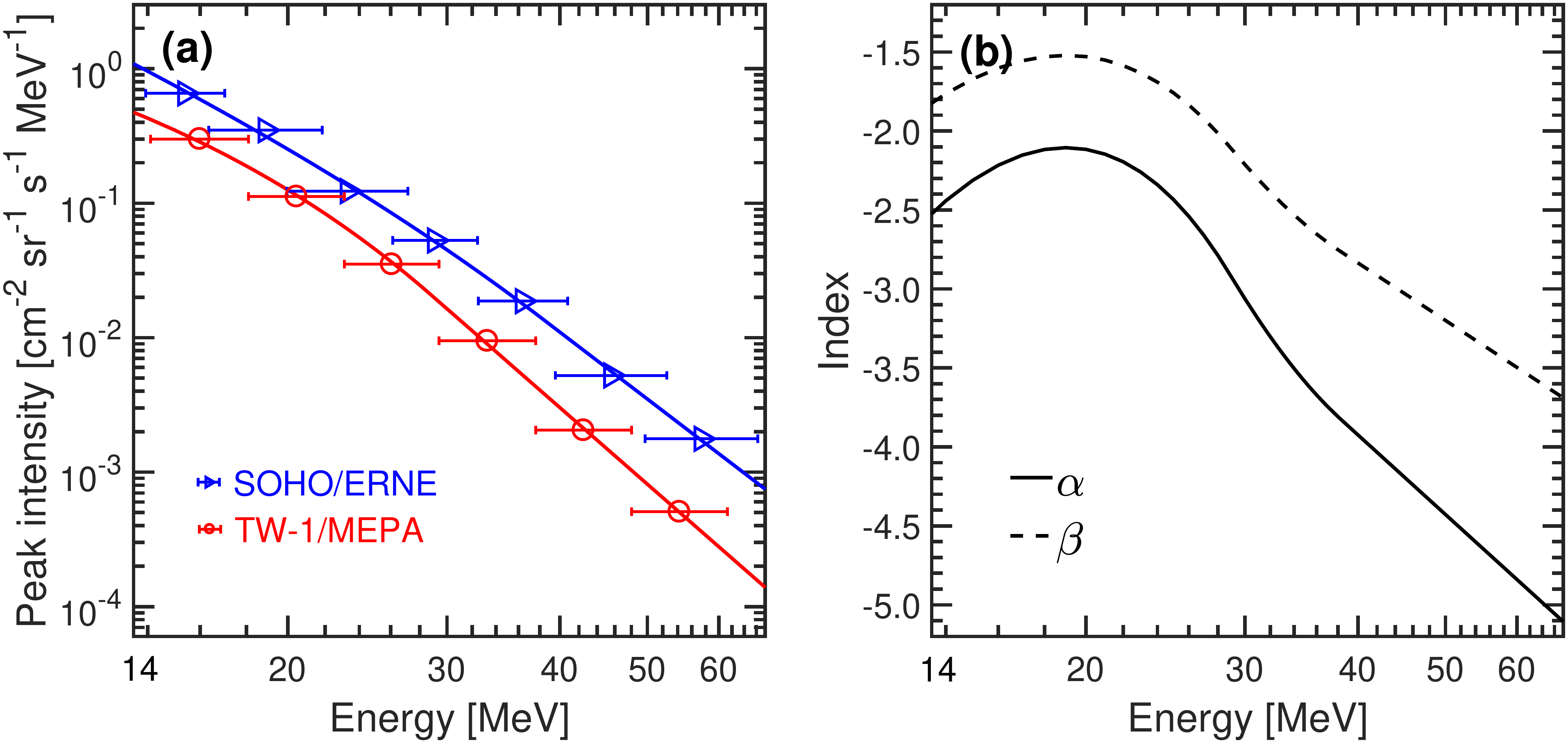}
   \caption{(a) Comparison of the SEP peak intensities between TW-1/MEPA and SOHO/ERNE (A). The scatter points are from spacecraft observations, and the curves are from a Band-function fit. (b) Calculated values of the index $\alpha$ (solid line) and $\beta$ (dashed line). See the text for details.}  \label{Fig4}
\end{figure}

The dependence of SEP peak intensities on radial distance and longitude has been examined in many previous works, such as \citet{lario2006radial,lario2013longitudinal,he2017propagation} and references therein. In these phenomenological studies, the radial dependence of peak intensities of SEPs is often fitted by $R^{\alpha}$ in Equation (\ref{eq2}). As discussed above, in this event, the Earth and TW-1 are connected in the same field line, and the effects of the inhomogeneous source distribution along the shock can be neglected. Therefore, this is an ideal case for studying how the peak intensity varies along a Parker field line and/or the radial dependence. Figure \ref{Fig4}(a) plots the peak intensities of SEP observed by SOHO/ERNE and TW-1/MEPA, respectively. Due to the differences in effective energies between the two instruments, we use a Band-function form in Equation (\ref{eq1}) to fit the observation data from 13.8 to 67.3 MeV. We then calculate the index $\alpha$ as a function of proton energies from the two fitted curves as shown in Figure \ref{Fig4}(a) (colored in blue and red), using Equation (\ref{eq2}). Besides, to study the IMF path length ($L$) dependence, we assume a $L^{\beta}$ relation for the SEP peak intensities, and obtain the index $\beta$ from Equation (\ref{eq3}). We plot the calculated indexes $\alpha$ and $\beta$ as a function of energies (13.8--67.3 MeV) in Figure \ref{Fig4}(b). Since the particles propagate along magnetic field lines, the IMF length dependence ($L^{\beta}$) is more physical and proper than the radial dependence.

From Figure \ref{Fig4}(b), we can see that although the values of $\alpha$ and $\beta$ are very different, they both have similar variation trends and significant energy dependence. \textcolor{black}{And as expected, the magnitude of $\beta$ is smaller than that of $\alpha$ due to the winding nature of the Parker spiral, which means that the path length increases much faster than linearly with radial distance.} Given that previous works, including both observations and numerical simulations, focused mainly on the study of $\alpha$, we will also discuss it in the following. As shown in the solid line of Figure \ref{Fig4}(b), the radial dependence index, $\alpha$, has a clear two-stage energy dependence, specifically, it increases with increasing energy below $\sim$20 MeV and varies between -2.5 to -2.0, whereas it decreases with increasing energy for energies above $\sim$20 MeV. In some early works (e.g., \citealp{smart2003comment,lario2006radial}, and references therein), proton fluxes from 1 AU to radial distances $>$ 1 AU are recommended to use a functional form of $R^{-3.3}$ with variations ranging from $R^{-4}$ to $R^{-3}$, based on both the results from a numerical model of diffusive energetic particle transport in the heliosphere and the spacecraft measurements at energies of 10 to 70 MeV from 1 to 5 AU. \textcolor{black}{Using the 1D Particle Acceleration and Transport in the Heliosphere (PATH) model \citep{Zank+etal+2000,Rice+etal+2003,li2003energetic}, \citet{ruzmaikin2005radial} found $\alpha$ in the range of -2.3 (low energy) to -2.9 (high energy).} In the work of \citet{lario2006radial}, they reported that $\alpha$ varies from -2.7 to -1.9 for 4--13 and 27--37 MeV proton peak intensities in a statistical study of 72 SEP events. However, this statistical work neglected the longitudinal separation of the spacecraft to the source location and the influence of the inhomogeneous source from the shock. In the modeling work of \citet{he2017propagation}, it was suggested that $\alpha$ $\sim$ -1.7 is the lower limit of SEP peak intensities in the case of observers at different radial distances but connected to an impulsive source along the same field line. \citet{he2017propagation} further suggested that this index does not strongly depend on the particle energies and the ratio of perpendicular diffusion to parallel diffusion, in contrast to the work of \citet{lario2007radial}, who suggested that $\alpha$ highly depends on the mean free path of the protons which can have a strong dependence on particle energy. \citet{lario2007radial} suggested that 1) a smaller mean free path leads to a larger decrease of peak intensity with radial distance, and 2) the higher particle energy, the smaller decrease of peak intensity with radial distance. In contrast, our results show that the higher the energy of protons, the larger decrease of peak intensity with larger radial distances. \textcolor{black}{Interestingly, such an energy dependence was predicted in the work of \citet{ruzmaikin2005radial} and was also confirmed in subsequent event modeling by \citet{Verkhoglyadova+etal+2009,Verkhoglyadova+etal+2010} using the PATH model.} Although it was a 1D model, the work of \citet{ruzmaikin2005radial} correctly captured the radial dependence of the peak intensity for this event. Note that in this event, the source of the accelerated protons is located east of the magnetic footpoints of the Earth and TW-1, and the particles propagate to the field lines that connect to the Earth and TW-1 in a longer duration, which amplifies the effects of the transport process, including both parallel diffusion and perpendicular diffusion, as well as adiabatic cooling. A better understanding of the radial dependence calls for a 2D/3D modeling effort, which can be achieved by the improved PATH (iPATH-2D) model \citep{hu2017modeling,hu2018modeling}.
%,ding2020modeling,hu2022extreme,li2021b
\section{Discussion and Conclusion}
\label{sect4}

The space radiation environment near Mars has attracted enormous attention, especially in recent years, as some government space agencies are planning crewed explorations of the Red Planet. Measurements from NASA$'$s MSL/RAD (Mars Science Laboratory/Radiation Assessment Detector) instrument showed that with the propulsion systems and shielding conditions at that time, the radiation dose equivalent was about (0.66 $\pm$ 0.12) sievert for even the shortest round trip \citep{zeitlin2013b}, implying a significant health risk from energetic particle radiation on any human mission to Mars. In contrast to chronic GCRs, sporadic and unpredictable SEP events tend to cause short-lived but extremely severe radiation effects that can endanger the lives of astronauts and the normal operation of electronic devices onboard the spacecraft. As an interesting case, during the SEP event in September 2017, the radiation level on the surface of Mars temporarily doubled compared to solar quiet periods \citep{hassler2018space}. Clearly, an accurate assessment of the space radiation conditions leading to and near Mars is a necessary prerequisite for designing a crewed Mars mission and for a long-term human presence on Mars.

In the SEP event on 29 November 2020, the Chinese TW-1/MEPA instrument happened to be switched on and measured the event from a radial distance of about 1.39 AU. This is the first SEP event encountered by TW-1 since its launch in July 2020. In this work, we performed a comparative analysis of energetic charged protons observed by TW-1/MEPA and near-Earth spacecraft, and discussed the potential implications of the observations on the underlying particle acceleration and propagation process in the heliosphere.

As the first widespread SEP event of the solar cycle 25, the 29 November 2020 event was observed simultaneously by multiple spacecraft within 1 AU, including PSP, STA, SolO, and near-Earth spacecraft (SOHO, ACE, and Wind). The TW-1/MEPA measurements complement these observations at a radial distance of $\sim$1.39 AU, providing additional help in revealing a complete picture of SEP propagation in the inner heliosphere. As discussed earlier, we assume that the solar wind speed between TW-1 and the Earth is the same in this event. And when the event occurred, the Earth and TW-1 were magnetically well connected since there was only a slight longitudinal separation ($\sim$5 degrees) between their corresponding magnetic footpoints at the Sun, which is the so-called Hohmann Parker effect. In fact, such a good magnetic connectivity can occur on all Hohmann transfers between adjacent planets in the inner heliosphere, e.g., between Earth and Mars, between Mercury and Venus, and between Venus and Earth, as discussed in the literature \citep{posner2013hohmann}.

In this work, we first compare the observed proton time-intensity profiles between the TW-1/MEPA and near-Earth spacecraft. The long onset delay at the Earth and TW-1 indicates that they did not connect to the shock via the nominal Parker field early on. As a result, cross-field diffusion of energetic particles is necessary to explain this event. Similar decay phases at TW-1 and Earth were observed, endorsing the reservoir phenomenon in the radial direction. Since both the Earth and TW-1 were not magnetically connected to the shock early on, we expect that the reservoir phenomenon also exists in longitude due to cross-field diffusion. We then compare the proton fluence spectra of Earth and TW-1. We found that these spectra are of broken power law shape and fit the spectra using a Band-function form. Not only the high-energy portion of the spectra at the Earth and at TW-1 are similar, the break energies are also similar. Since there is no radial dependence of the energy break and the high-energy spectral index, we suggest that the double power law feature of particle spectra is intrinsically caused by the acceleration process at the shock instead of due to particle transport process. Finally, we compute the radial dependence of peak intensities based on the observations of SOHO/ERNE and TW-1/MEPA in the energy range of 13.8--67.3 MeV. \textcolor{black}{Note that this energy range is above the spectral break point ($\sim$5.5 MeV).}
The obtained radial dependence index $\alpha$ shows a clear energy dependence and varies in a range of -5.5 $\sim$ -2.0, indicating a strong energy-dependent transport process where pitch-angle scattering, cross-field diffusion, and adiabatic deceleration all contribute. To further understand the reservoir phenomenon and the radial dependence of the peak intensities in this event, detailed numerical studies of this event using a model that addresses particle acceleration and transport in an integrated way need to be pursued.

This SEP event is only a moderate event with the maximum energy of the accelerated protons around 60 MeV. It is estimated that for atmospheric depths of $\sim$20 g/cm$^2$, only protons with an initial energy above 150 MeV can penetrate the Martian atmosphere and arrive at the surface (e.g., \citealp{guo2018generalized}). Therefore, these accelerated protons in this SEP event would be stopped in the Martian atmosphere and cannot trigger a noticeable variation in the radiation dose on the Martian surface. It should be remembered that as solar activity increases, stronger SEP events are inevitable in the future (especially during solar maximum years), which could significantly increase the acute radiation dose near/on Mars by several orders of magnitude. From this perspective, continuous monitoring of the energetic particle radiation environment around Mars will be very meaningful and essential. Furthermore, radiation exposures associated with highly energetic and highly penetrating GCR particles can be dominant at the Mars during solar quiet times. TW-1/MEPA is currently operating in a 265 km$\times$12,000 km elliptical orbit and continues to harvest more high-quality energetic charged particle data for the space science and engineering communities, which will potentially help design future crewed missions to Mars by providing key parameters on how much radiation shielding is required during a mission to Mars.

\acknowledgments
This work is supported by the Science and Technology Development Fund (FDCT) of Macau (Grant Nos. 020/2014/A1, 008/2017/AFJ, 0042/2018/A2, 0002/2019/APD, 0089/2018/A3, and 0049/2020/A1), the National Natural Science Foundation of China (NSFC) (Grant No. 11761161001), and the Pre-Research Project on Civil Aerospace Technologies of China National Space Administration (Grant No. D020101). S.F. acknowledges partial support from the Zhuhai Science and Technology Innovation Bureau. All authors are very grateful to all the data providers who provide valuable observation data to make this work possible.

\bibliography{my_ref}{}
\bibliographystyle{aasjournal}
\end{document}